\documentclass[twocolumn,pra,aps,superscriptaddress]{revtex4}

\usepackage{color}
\usepackage{epsfig}
\usepackage{latexsym}
\usepackage{amssymb}
\usepackage{amsmath}
\usepackage{algorithm}
\usepackage{algorithmic}
\usepackage{wrapfig}
\usepackage[apple mac]{inputenc}
\usepackage[english]{babel}
\usepackage{times}
\usepackage{latexsym}
\usepackage{fancyhdr}
\usepackage{verbatim}
\usepackage{tabularx}
\usepackage{epsfig}
\usepackage{amsmath}
\usepackage{amssymb}
\usepackage{graphicx}
\usepackage{wasysym}

\newcommand{\qed}{\hspace*{\fill}$\square$}

\newcommand{\be}{\begin{equation}}
\newcommand{\ee}{\end{equation}}
 \newcommand{\R}{\mathbf{R}}
 
 \newcommand{\N}{\mathbf{N}}

\newcommand{\fS}{\mathfrak S} %
\newcommand{\fX}{\mathfrak{X}} %
\newcommand{\fC}{\mathfrak{C}} %
\begin{document}

\title{Alan Turing and the Origins of Complexity}
\author{Miguel-Angel Martin-Delgado}
\affiliation{Departamento de F\i{}sica Te\'orica I, Universidad Complutense, 28040 Madrid, Spain}

% Journal of Scientific Exploration 23 (2009), pp. 223-225.

% Bulletin of the European Association for Theoretical Computer Science 97 (February 2009), pp. 157-164.

\begin{abstract} 
The 75th anniversary of Turing's seminal paper and his centennial year anniversary occur in 2011 and 2012, respectively.
It is natural to review and assess Turing's contributions in diverse fields in the light of new developments that his thoughts has triggered
in many scientific communities. Here, the main idea is to discuss how the work of Turing allows us to
change our views on the foundations of Mathematics, much like quantum
mechanics changed our conception of the world of Physics. Basic notions like computability and universality
are discussed in a broad context, making special emphasis on how the notion of complexity can be given
a precise meaning after Turing, i.e., not just qualitative but also quantitative. Turing's work is given some historical perspective
with respect to some of his precursors, contemporaries and mathematicians who took up his ideas farther.
\end{abstract}

\maketitle

\tableofcontents

%%%%%%%%%%%%%%%%%%%%%%%%%%%%%%%%%%%%%%
%%%%%%%%%%%%%%%%%%%%%%%%%%%%%%%%%%%%%%
\section{Introduction}
\label{sec:intro}
%%%%%%%%%%%%%%%%%%%%%%%%%%%%%%%%%%%%%%
%%%%%%%%%%%%%%%%%%%%%%%%%%%%%%%%%%%%%%

At the year of this writing, 2011, it is  75 years that  the seminal paper by Alan Mathison Turing \cite{turing_36} was published. And 2012 will mark his
centennial birthday.
It looks like a good occasion to review and assess his work and  impact. To See how useful and vital has become in so many aspects and disciplines.
As time goes by, the importance and relevance of this paper increasingly goes up. Turing transcends Mathematics and goes into other disciplines like Physics,
Engineering, etc. His work started as an in-depth and profound study of  the very notion of what an algorithm is, discarding irrelevant things and targeting the essence
of a mechanical procedure with a new notion of a theoretical machine. This so simple machine, the Turing machine, however turns out to be extremely powerful
and even universal. In this regard, Turing's work parallels Einstein's work on special relativity, when Einstein went to make precise and explicit definitions of 
elementary concepts like distances, time intervals, clock synchronization and the definition of an inertial frame. Despite their simplicity, however the consequences
of Einstein's principles revolutionized the whole Physics. Turing's work is of a similar kind.

Turing made a gigantic effort to understand how human's mind work at the level
of finding mechanical procedures to compute things and devise appropriate definitions of what algorithms are \cite{turing_36,turing_online}. 
His work represents  a great deal of imagination and creativity, which in turn has changed the notion of creativity ever since,
for creativity now can  be made quantitative using Turing's work.

He invented the theory of computability. What is more important, this affects the
way Mathematics must be understood at a fundamental level, the calculus.
And more. His results have revolutionized the way we should address axioms,
i.e., the very fundamentals of Mathematical disciplines. Questions like
when a set of axioms is complete or not, what to do when they are not complete.
Eventually, this leads to the very notion of mathematical creativity.

Turing got a lot of recognition in Engineering Schools, like Computer Science
and many others.
It is rather disappointing to see that the figure of Turing and his work still does
not have a central, pivotal role in the curricula of Mathematics university schools, which merits
a word or two. Firstly, Turing's theory can be regarded as the fundamental of 
what a calculation should be in Mathematics. It underlines all previous knowledge
on calculus and analysis in Mathematics in a way that it was implicit before Turing.
After Turing, it is systematized in a way that it becomes mechanical and algorithmic:
the holy grail of any theory. Secondly, it affects the way axioms have to be considered
in Mathematics. The big surprise is that Mathematics is not closed in the sense predicted
by David Hilbert, but it is an open system capable of increasing its amount of knowledge 
by adding new axioms to a discipline. 

And the same goes for Physics university schools, where the part of computer science in the curricula
is mostly reduced to learning manuals of software instead of the fundamentals of computation.
Manuals are changeable, version after version, but Turing's foundations on computability theory remain.

Turing's work has led to the development of 3 major disciplines:

\noindent {\bf Computability}: it studies which problems can be computed and which can not be computed. This goes to the very limits of what is knowable.

\noindent {\bf Complexity}: once a problem is computable, solvable, then we need to know how difficult is to compute it. This quantification can be made in different
ways giving rise to different notions of complexity: algorithmic complexity, computational complexity and others that will be considered in Sect. \ref{sec:not_compute}, \ref{sec:complexity}.

\noindent {\bf Universality}: the new paradigm is to use a TM, the basis of a real computer, in order to solve problems. Then, we need to know how general these 
machines, the Turing machines, can be. Does every problem or reduced set of them require a particular TM? This is another fundamental discovery, the notion of 
a Universal Turing Machine (UTM): a machine that can simulate the functioning of any other TM. Here it is important to study how many resources we need to create
such a UTM.

These disciplines are part of computer science as envisaged by Turing and extends to many branches of science like Physics, Mathematics, Engineering, etc. This extension
will increase with our better understanding of Nature and will apply to more descriptive sciences like Biology.

From Turing's work it is apparent that with a finite set of axioms it is not possible to cover Mathematics as a whole. There are irreducible truths, axioms that are not self-evident
in the sense of Euclid or Hilbert, and must be added to as independent axioms. This makes a TOE (Theory of Everything)  of Mathematics impossible.

%%%%%%%%%%%%%%%%%%%%%%%%%%%%%%%%%%%%%%
%%%%%%%%%%%%%%%%%%%%%%%%%%%%%%%%%%%%%%
\section{What you can compute...}
\label{sec:compute}
%%%%%%%%%%%%%%%%%%%%%%%%%%%%%%%%%%%%%%
%%%%%%%%%%%%%%%%%%%%%%%%%%%%%%%%%%%%%%

Turing was the first who separated software from hardware in a very concrete way. He did that by first focusing on the 
theoretical problem of having a well-defined notion of a computing machine. Later in his years, he also got involved in constructing 
computing machine in practice \cite{turing_calculator_45}.

Turing first goal was to scrutinize all steps that a person realizes during a calculation, like arithmetic, 
and separate irrelevant aspects from the relevant properties necessary to carry out the calculation.
In doing so, he realized that there were two relevant ingredients: 'local information' and 'state of mind'.
Local information means that at each calculation step, only a small part of the whole operation is being performed.
State of mind means that the steps after a local calculation is carried out, depend on the rules stored in the person's mind
which defines the calculation itself. Turing realized that it was enough to use a one-dimensional roll of paper or pad to 
write the intermediate (local) calculations and that the rules of the state of mind could be also stored in a table of operations.
After this analysis, Turing came up with a abstract construct of his machine.

\noindent {\bf Turing Machine}: it is a finite-state machine with 3 components:
i/ a doubly-infinite one-dimensional tape where symbols from an alphabet were written or read from square cells;
ii/ a control unit that stores the set of instructions in a table of specific operations;
iii/ a head that scans one cell of the tape at a time  and reads or writes alphabet symbols onto the tape depending on the instructions in
the control unit.
A more formal mathematical definition with the concrete functioning, examples and diagrams can be found in Galindo and Martin-Delgado \cite{rmp}.

He was so convinced that this definition of machine represented the most general possible algorithm for calculus that he formulated the basic principle of computation
by means of his construction:

\noindent {\bf Turing Hypothesis}: (also known as the Church-Turing thesis):

``A function is computable, if and only if, it can be computed by a Turing machine." 

Turing named his machine `a-machine' for automatic machine \cite{turing_online}.
In essence, this statement is more than a mathematical axiom, it is  part of Physics for it is a principle that tells us what we can compute in our Universe.

A basic and fundamental result of the notion of a TM is that the set of TMs is countable, infinitely denumerable. It corresponds to bit-strings. 
Let $\fX:=\{\Lambda, 0,1,00,01,10,11,000,\ldots\}$ be the set of finite strings of binary bits, with $\Lambda$ denoting the blank space symbol.
The size or number of bits is $|x|$.
The set of infinite bit-strings is denoted as $\fX^{\infty}$. A Turing Machine TM is an application $T: \fX \times \fX \rightarrow \fX$ that takes an input data $q\in \fX$ and a program
$p\in \fX$ that acts on the input to produce an output string $T(p,q)=x\in \fX$ which is the result of the computation, assuming it halts.  When the input data is empty, we simply write
$T(p)=x$, and when the output is simply
stopping the computer with no output, we write $T(p):\text{halts}$.

However, the notion of a TM is tight to the computation of a given function or problem. Changing the function means changing the TM. Here comes the notion
of universality as a property of a special TM that can compute what any other TM can do.

\noindent {\bf Universal Turing Machine}: denoted as UTM, it is construction based on set of instructions and states in the control unit of a TM such that
it can reproduce the functioning of any other TM.

It is very remarkable that the definition of a TM allows for this property of universality.
The basic idea behind the UTM is the observation that a TM  $T$ can be described by a bit-string itself and supplied to another TM $T^{\star}$ along with input data $q\in \fX$.
Thus,  $T^{\star}(T,q)$ will produce the same result as $T(q)$, thereby $T^{\star}$ simulating the functioning of any TM $T$.

In doing so, Turing was giving birth to programming and compiling.
A universal TM  is the notion of a general-purpose programmable computer of today.
After Turing gave the first construction of a UTM \cite{turing_36}, other constructions have been presented depending on the number of states used by the machine
and the number of symbols in the alphabet \cite{shannon_56,minsky_62}, including small ones \cite{Rogozhin_96}.

Von Neumann realized that Turing had achieved the goal of defining the notion of universal computing machine,  and went on to 
think about practical implementations of this theoretical computer. It was clear that this was the crucial notion
of a flexible computer that was needed and was lacking thus far. Therefore, the distinction between software and
hardware is clear in Turing's work and it is a consequence of it. Turing did not care about practical implementations
at his time because he wanted to isolate, to single out the very notion of what a computer is, in theory. In doing so,
he was inspired by D. HIlbert and his ideas about a formal set of axioms from which theorems would be provable by means
of a mechanical procedure. This led to the notion of TM and the solution of Hilbert's tenth problem.

As the title of his 1936 paper states, Turing wanted to give a concrete definition of what a computable real number is. By introducing the TM, he identified
computable numbers with those that a TM can really compute. Thus, a real number is computable when its decimal digits are computable by finite means.

\noindent {\bf Computable Numbers}: a real number $x\in\R$ is a computable real if there exists a computable function $T(k)$, $k\in\N$ such that $x$ is bounded
by rational numbers:
\begin{equation}
\frac{k-1}{n} < x < \frac{k+1}{n}, \ \forall n\in\N.
\end{equation}
Fortunately, all algebraic numbers, as well as, $\pi$, ${\rm e}$, and many other transcendental numbers are computable reals. 

In addressing non-computable problems in Sect.\ref{sec:not_compute}, it is useful to introduce a variant of Turing machine due to Chaitin \cite{chaitin_75,chaitin_87} .

\noindent {\bf Chaitin Machine}: it is a self-delimiting or prefix-free Turing Machine, denoted CM.

This means that the TM knows when to stop by itself, without needing an special mark indicator or blank character.
Formally, it is an application $C: \fX \times \fX \rightarrow \fX$ that is a TM acting on programs $p\in \fX$ and input data $q\in \fX$,
such that both $p$ and $q$ are self-delimiting strings, also called prefix-free. A set of strings $\fS\subset \fX$ is prefix-free if $\forall s,s'\in \fS$,
$s$ is not included as a prefix in $s'$.  For example, the set of all bit strings up to size 2,  $\fX_2:=\{0,1,01,10\}$ is not prefix-free for $0$ is prefix of $01$.
However,  $\fS=\{0,10\}$ is prefix-free.

An explicit construction of a Chaitin machine is as follows. 
It has three elements: i/ a finite program tape; ii/ a doubly-infinity work tape; iii/ a head with one arrow scanning the program tape and another arrow scanning the work tape.
The alphabet is binary   $0,1$ and the blank space is not allowed to mark the halting of the machine.  The initial state of a CM is the program $p\in\fX$ stored in the program
tape and with the arrow head scanning the left-most square which is blank. As for the work tape, it is occupied with the input data $q\in\fX$ and the arrow head is scanning
the left-most bit (initial bit) of $q$. After the initial state, the CM starts operating like a TM: the arrow head only moves on the program tape to the right, while the arrow head
can move left/right on the work tape; the arrow head can read and erase the square of the work tape being scanned. The CM will halt if the arrow head
reaches the right-most square of the program tape, giving a certain output result $C(p,q)=:x\in\fX$; otherwise, $C(p,q)$ is not defined and does not halt.
Exactly as with  ordinary Turing machines, the CM moves step by step  following a previously
given finite table that completely determines the computation for the
argument $(p,q)$.

Notice that this construction of a TM is self-delimiting since the read arrow head cannot read-off the right-most square of the finite program tape.  Also, in an ordinary TM,
a program that halts is necessarily prefix-free: it cannot be extended into another program that halts.

There exists procedures to make a given set of bit-strings into a self-delimiting set. For a bit-string $x$ we construct a new bit-string by appending to it a prefix depending on
its length $|x|=:n$ as follows:
\begin{equation}
x_s:= 0^n 1 x.
\end{equation}
For instance, from the above $\fX_2$ we construct $\fX^s_2=\{010,011,00101,00110\}\subset \fX_4$, which is prefix-free.
Thus, the length increases only by an additive logarithmic term in the
transition from a bit-string to its self-delimiting presentation:
\begin{equation}
|x_s| = |x| + 2\log|x|,
\end{equation}
asymptotically.
An important property is that universal Chaitin Machines also exist: the universal CM $U$ starts reading a prefix-free program $\pi_C$ that indicates which CM to simulate,
followed by the binary program for that machine, $U(\pi_C p)=C(p)$, with $p$ also prefix-free. The whole input program for $U$ can also be made prefix-free.

%%%%%%%%%%%%%%%%%%%%%%%%%%%%%%%%%%%%%%
%%%%%%%%%%%%%%%%%%%%%%%%%%%%%%%%%%%%%%
\section{... And what you can not compute}
\label{sec:not_compute}
%%%%%%%%%%%%%%%%%%%%%%%%%%%%%%%%%%%%%%
%%%%%%%%%%%%%%%%%%%%%%%%%%%%%%%%%%%%%%

It is a twist of destiny that in the same paper where Turing shows what we can compute
in a very precise and universal way ... he also proves that there are things that we cannot compute.

G\"{o}del's theorem on incompleteness \cite{godel_31} was a first shock for the foundations
of Mathematics as a complete formal logical system. The latter was the attitude
predominant before and well represented by David Hilbert. Yet, the real impact
of G\"{o}del's was still under debate in the Mathematics community and there was
the impression that they were a kind of minor anomaly that would not affect the
whole building of the theory.
Turing's non-computability 
results were even more demolishing for the fundamentals of Mathematics since
he showed that a very important example of G\"{o}del's results was also at the 
heart of computation, algorithmic, something very practical and with a lot of impact
in the future.

It is easy to write programs, in pseudocode, that will never halt:
\begin{equation}
\text{while true, continue}
\end{equation}
will loop forever. Another less evident example of looping program is:
\begin{equation}
\begin{matrix}
\text{define}  & n \quad \text{integer number}; \\
\text{let}  \  n=1,&  \text{then}   \quad n = \begin{cases}
                                       \frac{n}{2} & \text{if}  \ n\equiv 0 \ \text{(mod 2)},\\
                                       3n+1         &  \text{if}  \ n\equiv 1 \ \text{(mod 2)}.
                                      \end{cases}
\end{matrix}
\end{equation}
It produces the cycle $1,4,2,1$ forever.

Thus, a skillful debugger may envisage the task of finding all possible loops in programs
and with a look-up table, to get rid of them. Or maybe, one has to study more and it is necessary to classify families of loops etc. Turing's proof
shows that this dream is impossible and does not depend on how smart the debugger is. It is at the roots of computational theory.
In fact, we can guess that the purposed debugger may easily run into unknown territory. For instance, we can use the {\em Collatz 
conjecture} \cite{collatz_37} to write the following simple program:
\begin{equation}
\begin{matrix}
\text{define}  & n \quad \text{integer number}; \\
\text{if}   \quad n=1,&  \text{stop};\\
\text{while}  \  n\neq 1,&  \text{let}   \quad n = \begin{cases}
                                       \frac{n}{2} & \text{if}  \ n\equiv 0 \ \text{(mod 2)},\\
                                       3n+1         &  \text{if}  \ n\equiv 1 \ \text{(mod 2)}.
                                      \end{cases}
\end{matrix}
\label{collatz}
\end{equation}
It has been checked that this program stops for very large values,  $n\leq 20\times 2^{58}$ \cite{numerical_collatz}, but it is unknown whether it halts $\forall n\in \N$. The conjecture remains unproven. A modification of it can has been proved to be undecidable \cite{kurtz_simon_07}, but the modification does not apply to the original conjecture.

A basic and fundamental result of the notion of a TM is that the set of TMs is countable, infinitely denumerable. It corresponds to bit-strings $\fX$. 
This is the power of TMs ... and also its weakness. Although we know that its cardinality is infinity, after Cantor we know that not all infinities are alike.
In particular, $|\fX|$ is an infinity equal to the infinity of the real numbers $\N$. This is easily obtained by seeing a bit-string as the binary representation of an integer number
in base 2: $n=\sum_{n=0}^{\infty} x_n 2^{n}$.

Cantor's diagonal method provides a clever  way to see that there
are more real numbers $\R$ than natural numbers $\N$ \cite{cantor_91,suppes_60,dauben_79}.

\noindent {\bf Cantor's Diagonal Method}: it is a technique in set theory to create a new element which is not an element of a previously given set of elements.

As an illustration, consider the following table where we place eight bit-strings $\fS:=\{x_1,x_2,\ldots,x_8\}\subset \fX$.
From this, we can construct another element $x_9\notin \fS$: select the diagonal of the table and negate each of its bits.
Then, we get $x_9:=00000000$ which is new.

\begin{equation}
\begin{array}{|c||c|c|c|c|c|c|c|c|} 
\hline
x_1 &   \not 1 & 0  & 0  & 0  & 0 & 0  & 0  &0 \\
\hline
x_2 &  0 & \not  1  &1   &  0 & 0 &  0 &  0 & 0\\
\hline
x_3 &   0& 0 & \not  1  & 0  & 0 & 0  & 0  & 0\\
\hline
x_4 &  0 & 0 &   0& \not 1   &1  &  0 &  0 & 0\\
\hline
x_5 &   0& 0 & 0  & 0  &  \not 1 & 0  & 0  & 0\\
\hline
x_6 &  0 & 0 &  0 &  0 & 0 &  \not 1  & 1  & 0\\
\hline
x_7 &   0& 0 & 0  & 0  & 0 & 0  &   \not 1  & 0\\
\hline
x_8 & 1  & 0 &  0 &  0 & 0 &  0 &  0 &  \not 1\\
\hline
\end{array}
\label{table}
\end{equation}

The diagonal method is very general. It applies both to finite sets like $\fX$, or infinite sets like $\fX^\infty$: the set of infinite bynary strings.
A consequence of this is that $\fX^\infty$ has infinite cardinality but it is uncountable. To show this, we proof it by reductio ad absurdum. Assume that 
$\fX^\infty$ is countable so that we make a table like \eqref{table} with infinite elements ordered by the integers $\N$. All elements are thus listed,
but with the diagonal we can create another bit-string $x_{d}$:
\begin{equation}
x_{d}:=\left( x_{i,i}\oplus 1 \right)_{i=1}^\infty ,
\end{equation}
where $ x_{i,i}$ is the ith bit of the ith listed element of $\fX^\infty$. But then, $x_{d}\notin \fX^\infty$, which is a contradiction. The assumption that
$\fX^\infty$ was a countable set is not true.

In fact, we can go on and prove that the set of real numbers  $\R$ is uncountable by establishing a bijection between  $\fX^\infty$ and $\R$. Both  $\N$ and $\R$  are infinite sets, but of a different quality.
The cardinality of  $\N$ is denoted by $\aleph_0$. $\R$ has the cardinality of the continuum. 

\noindent {\bf Continuum Hypothesis (CH)}: it states that the cardinality of  $\R$ is $\aleph_1$, the second transfinite cardinal introduced by Cantor, 
or equivalently, that every infinite subset of $\R$ must apply  bijectively on either $\N$ or on $\R$ itself: $2^{\aleph_0}=\aleph_1$.

In other words, there is no set with an intermediate
cardinality between $\N$ and $\R$, there is a gap.
CH was introduced by Cantor but was unable to prove it \cite{dauben_79}. It is the first of Hilbert's twenty-three problems proposed in 1900.
 G\"{o}del proved that CH is consistent with axiomatic set theory \cite{godel_40}, 
but Paul J. Cohen also proved that the negation
of CH is also consistent with the axioms of set theory \cite{cohen_63}. 
Thus, CH is undecidable or non-computable. It is independent of standard axiomatic set theory (Zermelo–Fraenkel set theory).

\noindent {\bf Non-Computable Numbers}: a real number non-computable by a TM.

The set of computable reals with a TM is quite small. Given a TM with $|S|$ internal states, it can compute about $(4|S|+4)^{2|S|}$ different numbers.
Using Cantor's diagonal method, Turing was able to prove that there are uncountably many noncomputable numbers. Most of the real numbers, the continuum, is
unaccessible to a TM.

The diagonal method has proved extremely useful in fundamental problems  of Mathematics. Some instances are Russell's paradox in set theory, G\"{o}del's first theorem
of incompleteness and Turing's solution to the 10th Hilbert's problem.

\noindent {\bf Halting Problem}: There is no way to find whether a computer will eventually halt.

A crucial assumption in Turing's formulation of this problem is that there is no limit for the running time of the computer.
By computer is meant a TM. Under these circumstances, there is no mechanical procedure that can decide in advance
whether a computer will ever halt.
A more formal statement is the following:

Let $H$ be the set of subsets, such that each subset corresponds to a  Turing Machines $T_n$, $n\in \N$ and all its programs that halt when input on $T_n$. Each
program can also be labeled with an integer $m\in \N$. Thus, the allegedly  total halting set is
\begin{equation}
H:=\{ (n,m)\in \N\times \N: T_n(m) \ \text{halts}\} .
\end{equation}
In bit-string notation $T_n(m):=T_{r_n}(p_m,0)$,i.e., input data $q=0$, the program $p_m\in\fX$ is the bit-string of the natural number $m$ and similarly for the bit-string
$x_r$ labeling TMs.  Each subset of $H$ is the halting set of a TM $H_n$:
\begin{equation}
H_n:=\{ m\in  \N: T_n(m) \ \text{halts}\} .
\end{equation}
Now, we are in the situation of applying Cantor's diagonal method. The set $H$ can be arranged as a table \eqref{table}, with $H_n$ being the rows. 
Let us define a 'diagonal' set $D$ as follows:
\begin{equation}
D:=\{ n\in  \N: n\notin H_n \} .
\end{equation}
By construction, $D$ is a set of natural numbers that is different from any halting set  $H_n$ of any TM. 
Therefore, the original goal of determining the set $H$ of all halting machines
cannot be accomplished and thus, we can never know in general when a TM will ever halt.

Turing did not use the terminology of `halting problem' in his 1936 paper \cite{turing_36,turing_online}. It seems that
the first time this was used was by Martin Davis \cite{davis_65,copeland_04}.

After Turing found an explicit and crucial example of a non-computable problem, it was natural to ask wether more examples of this kind could be found.
In 1962, T. Rad\'o \cite{rado_62,rmp} proposed another interesting non-computable function.

\noindent {\bf Busy Beaver Function}: it is the maximum number of digits 1s that appear as output $x$ in a TM T that runs over all programs $p$ that halt on no input 
$q=\Lambda$:
\begin{equation}
\Sigma_T:=\underset{p: T(p)=x}{\text{max} \  |x(1)|},
\end{equation}
wher $|x(1)|$ is the number of 1s in $x\in\fX$.
There are several variants of Busy Beaver functions that have the same property of being non-computable and are  more manageable definitions.
For instance, as the maximum integer number that can be named with a universal TM $U$ with programs of a given size $|p|=:N$.
Thus, a $N$-th Busy Beaver function is denoted $\Sigma_N$ and defined
\begin{equation}
\Sigma_N:= \underset{p: |p|\leq N, U(p)=k}{\text{max}\  \ k}.
\label{busy_beaver}
\end{equation}
This is a well-defined function $\Sigma_N: \N \rightarrow \N$, but it is noncomputable: it grows faster than any computable function $f(N)$, $\Sigma_N > f(N)$ for sufficiently 
large $N$. Therefore, $\Sigma_N$ cannot be bounded in the form of $\Sigma_N = O(f(N))$. The proof goes by reductio ad adsurdum: if it could be bounded, then the halting
problem would be computable. More examples of non-computable functions can be obtained systematically by means of the Algorithmic Information Theory (AIT) 
\cite{solomonoff_64, kolmogorov_65, chaitin_66}.

After Turing's halting problem, we may ask: can we quantify non-computability on mathematical grounds? G. Chaitin has done a great deal of work 
\cite{chaitin_75, chaitin_87,chaitin_01,chaitin_03} by approaching this
issue from information-theoretical methods. He has developed the concept of what it known as Chaitin's $\Omega$ number that allows us to address this fundamental question.
Thus, we need to introduce some basic concepts and results from AIT.

\noindent {\bf Algorithmic Information Theory (AIT)}: it is a part of Information Theory that deals with the algorithmic complexity of functions and problems. 
The algorithmic complexity of a program $p\in\fX$ refers to its program-size,i.e., bits of information regardless the run-time that a machine like a TM takes to execute it.
It is defined as the shortest program that can reproduce a given string $x$ in a universal
TM:
\begin{equation}
H(x):= \underset{p: U(p)=x}{\text{min} \ \  |p|}.
\label{complexity}
\end{equation}
A first consequence of this definition is that $H(x)$ is not computable itself, for two reasons: due to the halting problem, we never know when the programs will halt and
since it is a minimization procedure. Moreover,
it is not possible to compute lower bounds to $H(x)$. What is possible is to give upper bounds. These are good enough to gain a great deal of 
insight into a given problem. For instance, we can give an alternative definition to the Busy Beaver function: 
\begin{equation}
\Sigma_N:= \underset{H(k)\leq N}{\text{max}\  \ k},
\label{busy_beaver}
\end{equation}
where the algorithmic complexity \eqref{complexity} is defined for programs $p$ that compute $k=U(p)$ without input and halting.

Although non-computable, algorithmic complexity is well-defined and it has very useful properties like subadditivity: the joint complexity is bounded by the sum of the complexities
of the individuals:
\begin{equation}
H(x,y) \leq H(x) + H(y) + O(1).
\label{subadditivity}
\end{equation}
This allows us to construct big programs out of small ones.
Another crucial property follows from a proper definition of relative entropy $H(y|x)$
\begin{equation}
H(x, y) = H(x) + H(y|x^{\ast}) + O(1).
\end{equation}
Thus, the joint complexity of two bit-strings can be computed knowing the absolute complexity of the first one plus the relative complexity
of the second given the first one. The key point for this result to hold true is the definition of relative complexity of $y$ given $x$, $H(y|x^{\ast})$:
the size in bits of the smallest self-delimiting program for calculating $y$ if
we are given for free, not $x$ directly, but  $x^{\ast}$, a minimum-size self-delimiting
program for $x$.

A fundamental property of Chaitin machines is that they allows us to define halting probabilities for TMs, or the algorithmic probability of a bit-string,
also known as universal probability $P_U(x)$ of a given string $x\in \fX$:
\begin{equation}
P_U(x):=\sum_{p:U(p)=x} 2^{-|p|},
\label{universal}
\end{equation}
which is the probability that a program randomly drawn as a sequence of
fair coin flips $p=p_1 p_2\ldots$ will compute the string $x$. 
This is well-defined thanks to the prefix-free property of CMs and results from AIT \cite{chaitin_75, chaitin_87,chaitin_01,chaitin_03}.

A central theorem relates algorithmic complexities with algorithmic probabilities:
\begin{equation}
H(x) = -\log P_U(x) + O(1).
\label{relation}
\end{equation}
This relation tells us that near-minimum
size programs for calculating something, elegant programs, are essentially unique.
This is a mathematical formulation of Occam's Razor. Essentially, this relation tells us
that AIT is equivalent to Probability Theory, although this probability has to do with 
randomness in programs, rather than statistical randomness but we shall get back to this later.

 The idea behind structural or logical randomness is lack of structure or pattern
in a program or bit-string. Thus, a program or bit-string is random if it has no pattern or inner structure,
consequently, it cannot be compressed. The only way to address it is by printing the whole program as it is:
there is no theory behind it from which it can be derived. By theory, we mean a simpler procedure to recover the bit-string,
something compressible. 
Now, we can give a precise definition of randomness using information-theoretic notions like
algorithmic complexity. This was definded by Chaitin in AIT. It is necessary to distinguish between finite bit-strings $x\in\fX$ ($|x|=:n<\infty$, and infinite bit-strings $x\in\fX^{\infty}$, $x=(x_n)_{n=1}^\infty$
\cite{chaitin_01}:

\noindent i/ Random finite bit-strings:
\begin{equation}
H(x) \approx n + H(n).
\label{randomness1}
\end{equation}
\noindent ii/ Random infinite bit-strings:
\begin{equation}
H(x_n) >  n - c, \ c=\text{const.}, \forall n.
\label{randomness2}
\end{equation}
Notice that $ n + H(n)$ is the greatest possible and also typical complexity of a finite bit-string. Equivalently, the relative complexity $H(x|n)\approx n$.
As for infinite bit-strings, it is required that the partial series of bit-strings $x_n$ always be as random as possible.

It is possible to prove that the definition of randomness for infinite strings from AIT \eqref{randomness2} is equivalent to the statistical definition of random real numbers
in classical probabilistic theory 
introduced by Martin-L\"{o}f \cite{martin_lof_66} and Solovay \cite{chaitin_01}. This is a very remarkable result since the origin of AIT randomness is conceptually different and related to lack of logical structure in a set of programs. It is very nice that both types of definitions produce exactly the same infinite random sequences 
\cite{li_vitanyi_90,calude_02,calude_et_al_02}.
Moreover, for finite bit-strings AIT also provides a definition of randomness.

We can now define Chaitin's $\Omega$ number and use it to assess logical randomness in 
Information Theory, the issue of non-computability. The motivation is to define the halting 
probability of a TM, i.e., 
\begin{equation}
\Omega:=\sum_x P_U(x),
\end{equation}
where the sum runs over prefix-free strings and the universal computer $U$ is a Chaitin machine.
This way, $\Omega$ can be thought of as an average on the Turing halting problem.
It is possible to give a more explicit expression as follows:
\begin{equation}
\Omega:=\sum_{p:U(p)=\text{halts}} 2^{-|p|},
\label{omega}
\end{equation}
 It measures the probability that a randomly chosen program $p$ will halt when run in a universal TM $U$
that halts. This follows from the definition of $P_U(x)$. It is a well-defined probability for: i/ only self-delimiting programs are allowed; ii/ thus, the sum is convergent
due to the Kraft inequality  \cite{cover_thomas_06}; iii/ $0<\Omega<1$, because there are always programs that halt and also programs that never halt.
Alternatively, we can use algorithmic complexity to define it:
\begin{equation}
\Omega:=\sum_{x} 2^{-H(x)}.
\label{omega}
\end{equation}
What is behind $\Omega$ is a very compact way of encoding the halting problem, or any other non-computable problem.

The Chaitin $\Omega$ number is a real number in $(0,1)$ which is logically random \eqref{randomness2}: let us truncate it up to programs of bit-size $N$,
\begin{equation}
\Omega_N:=\sum_{p: |p|<N} 2^{-|p|}.
\label{omega_N}
\end{equation}
then, it is possible to prove that $H(\Omega_N)>N-c$, $\forall N$ and certain constant $c$. $\Omega$ is algorithmically random and incompressible.
These $\Omega_N$ are lower bounds to the actual $\Omega$.
This truncation also produces an unbounded function $\Omega_N$ that reflects its non-computability.
Knowing the first $N$ bits of $\Omega$, i.e., the binary expansion of $\Omega_N:=0.\omega_1 \omega_2 \ldots \omega_N$ then it is possible
to decide the truth of $N$-bit theorems. By construction, knowing $\Omega_N$  enables us to decide  all programs
of length $|p|<N$ that halt. Now, for instance, it is possible to write a program that searches for solution of the Collatz conjecture \eqref{collatz} and halts only if a 
counterexample is found. Knowing sufficiently long string bits of $\Omega$ enables us to decide whether a well-defined problem, according to a formal theory,  is a
theorem, a non-theorem or independent.

After having faced the limits of computability, the natural question is: can we go beyond? This depends on what is called the Turing barrier
\cite{feynman_82,berstein_vazirani_97,calude_pavlolv_02,kieu_03}, that is stated as follows.

\noindent {\bf Turing Barrier}:  there is  no way whatsoever to beat the halting problem.

This notion has originated a line or resesearch called Hypercomputation. It speculates that it is possible to devise theoretical or physical machines that can compute
problems that are non-computable by the TM model \cite{copeland_et_al_99}.

%%%%%%%%%%%%%%%%%%%%%%%%%%%%%%%%%%%%%%
%%%%%%%%%%%%%%%%%%%%%%%%%%%%%%%%%%%%%%
\section{Precursors of Turing on Computability}
\label{sec:precursors}
%%%%%%%%%%%%%%%%%%%%%%%%%%%%%%%%%%%%%%
%%%%%%%%%%%%%%%%%%%%%%%%%%%%%%%%%%%%%%

The following list  has, by no means, the intention of giving a full account of all who
might have been involved  directly or indirectly on investigations touching upon Turing's work,
but simply to present some important facts that are interesting in connection to his work and later developments.
Due to space constraints we cannot dwell upon the work of such as 
Georg Cantor (the diagonal  method \cite{cantor_91}, cardinalities \cite{dauben_79}), David Hilbert (the axiomatic method \cite{hilbert_27}), \'Emile Borel (normal sequences \cite{borel_09}, the inaccessible number \cite{borel_52}) etc. that nevertheless will appear mentioned along the way.

\subsection{Gottfried W. Leibniz}

Leibniz made a crucial discovery that today is taken for granted but is 
a major breakthrough in computational theory: the binary numeral system (base 2)
$\{0,1\}$ as a system for calculus. He went on and fabricated a mechanical machine
that worked simple multiplication operations with this binary system.
He dreamt of human reason reduced to calculation and of 
powerful mechanical engines to carry out  those calculations. 

Leibniz asked and thought about fundamental questions and ideas about what is Science
and Nature \cite{chaitin_09}. They  play a central role in modern scientific methodology. One of these questions
he asked was:  is there any difference between a world without laws of nature and a world described by
laws? How can we tell the difference. Today this looks pretty obvious after the enormous success of the
scientific method for about more than three hundred years. But Leibniz analysis was made in 1686 \cite{leibniz_86} (another celebration in this year 2011)
one year before the Newton's Principia were published \cite{newton_87}. The mechanistic view of the world was not predominant whatsoever.

In addressing those questions, Leibniz touched upon the roots of what a physical law must be: simplicity must be the key.
To show this, he posed a very concrete mathematical example. Suppose you are given a set of points in a plane that they
represent the experimental data you want to explain by a law. 
It is well-known from interpolation techniques, like Lagrangian interpolation he anticipated, that we can always find a function that fits a given finite number of points.
How do we know then, that a physical law exists behind them? Leibniz's answer is:  only if the rule to fit the data is simple enough. 
His basic principle is Occam's Razor. 
With Turing, we know how to quantify complexity for instance by means of the notion of compression.

Leibniz also stated that the Universe has a duality relationship between complexity vs. simplicity.
On one side, Universe is extremely diverse and rich, complex. On the other hand, it can be made out of
very simple rules that we call fundamental laws. Complexity out of simplicity, like in a Beethoven's symphony.
In the computer's era of today, we have a typical 
example of this phenomenon: a laptop computer can produce a fabulous number of complicated images, movies, games etc.
Yet, all there is underneath is Leibniz's binary system. 
In this way, he anticipated the notion of emergent phenomena that is so influential and modern  in theoretical physics.

\subsection{Hermann K.H. Weyl}

Weyl became interested in Mathematical Logic and the foundations of Mathematics since his thesis supervisor was David Hilbert in Gottingen.
He wrote a thorough book \cite{weyl_49} on these topics in which he calls the attention of Leibniz's unpublished work \cite{leibniz_86} on the nature of a physical law
and science. Weyl discussed on the character of mathematical cognition, the axiomatic method and natural science.

He declares that the problem of simplicity is of central importance for the
epistemology of the natural sciences. As an example of the principle of
simplicity in physics, he claims that  it is a sure sign of being on the wrong scent if one's theory
suffers the fate of the epicycles of Ptolemy whose number had to be
increased every time the accuracy of observation improved. The
three laws of Kepler were much simpler and yet agreed noticeably
better with the observations than the most complicated system of
epicycles that had been dreamed up. 

Weyl took Leibniz's thoughts about complexity to the extreme case and established that if we allow arbitrary high complexity
in a law of physics, then the law ceases to be a law ... because then there is always a law.  Thus, some sort of balance has to be reached.

Admitting that  the concept of simplicity
appears to be so inaccessible to objective formulation, he failed to come up with a precise definition of complexity, see Sect.\ref{sec:complexity}.

\subsection{Kurt F. G\"{o}del}

In year 1931 G\"{o}del surprised the great mathematicians of his time by showing that Hilbert's proposal of finding a complete axiomatic formalization
of Mathematics was impossible \cite{godel_31}. This was shocking since it was  like  if the ultimate goal of Mathematics, its reason of being, could not be achieved. Von Neumann
was the first to realize that G\"{o}del was correct even before his publication by attending a conference by G\"{o}del in K\"{o}nisberg. Subsequently, Weyl and others
had to concede as well that he was right. G\"{o}del was a great admirer of Leibniz and studied his works thoroughly.

A common misconception about G\"{o}del's work is that it is destructive towards Mathematics since it looks like
an attack at what Mathematics was understood to be: a well-defined formal system to solve problems. Quite on the contrary,
this objective is still true after G\"{o}del's results, but has to be revised and made precise by considering incompleteness as a key
ingredient in Mathematics. Although people think that G\"{o}del's theorem are bad news, a closer analysis reveals that they are
good news and positive results since it allows creativity to become a key role in the foundations of Mathematics and this can be
done in a rigorous way as it demands.

The heart of G\"{o}del's proof relies is using a self-reference proposition like 
\begin{equation}
\text{'This statement is unprovable'} 
\label{liar1}
\end{equation}
or equivalently, the liar's paradox
\begin{equation}
\text{'This statement is false' or 'I'm lying'}, 
\label{liar2}
\end{equation}
to undermine the logical system of Hilbert and followers. 
The latter was based on a set of axioms from which the proof of theorems followed like a mechanical checker. 
Whichever option you take on the statement \eqref{liar1},\eqref{liar2}, true or false, you get the opposite.
Then, G\"{o}del went on performing a series of transformations
into that initial paradox, some of them involving properties of prime numbers,  and making it into definite statements in number theory.
And this was very clever and imaginative. As such, one cannot ignore a statement in number theory which is not provable. 
Hence, G\"{o}del's results deserved to be taken seriously.

In year 1936 Turing gave a second and definitive surprise to the community of mathematicians by proving the existence of non-computable problems, providing
an explicit example. His result can be seen as an instance of G\"{o}del's result, but much simpler to understand and, at the same time, playing a central role in the theory
of computation.
% In fact, G\"{o}del's incompleteness theorem can be given a version in terms of TMs %%% Check this

When time gives more perspective to   G\"{o}del's work, it will be considered similarly to what happened with the advent of non-Ecludian geometry
in the XIX century, or more plainly, how the discovery of irrational numbers shocked the Pythagorean dreams.

%%%%%%%%%%%%%%%%%%%%%%%%%%%%%%%%%%%%%%
%%%%%%%%%%%%%%%%%%%%%%%%%%%%%%%%%%%%%%
\section{Computability after Turing}
\label{sec:successors}
%%%%%%%%%%%%%%%%%%%%%%%%%%%%%%%%%%%%%%
%%%%%%%%%%%%%%%%%%%%%%%%%%%%%%%%%%%%%%

The same applies for the farther developers of Turing's theory  as with his predecessors, and with the
same proviso on the number of figures that should be mentioned. For instance, all the recipients of the Turing award \cite{turing_award}.

\subsection{Tibor Rad\'o}
Rad\'o made a great contribution in the theory of Turing Machines in his late life 1962, three years before his death and after having
accomplished major contributions in other fields of Mathematics: he solved the plateau problem, discovered essentially unique triangulations
of surfaces,  and made many other important contributions in conformal mappings, real analysis, calculus of variations, subharmonic functions, 
potential theory, partial differential equations, integration theory, differential geometry, and topology.

He invented the Busy Beaver function \cite{rado_62}, another example of non-computable
function after Turing \eqref{busy_beaver}.

\subsection{Gregory J. Chaitin}

Gregory J. Chaitin, together with Ray Solomonoff and Andrei N.
Kolmogorov, are the founding fathers of the subject called Algorithmic
Complexity, Kolmogorov Complexity, or Algorithmic Information Theory
(AIT) \cite{solomonoff_64, kolmogorov_65, chaitin_66}.

Chaitin approached the two fundamental discoveries by G\"{o}del 1931 and Turing 1936 and his 
assessment was that they were just the tip of the iceberg. Those were not isolated marginal results, but
they were the natural case in Mathematics rather than the exception. Those results implied that in some parts
of Mathematics, it was possible to have lack of structure, of patterns, a sort of randomness intrinsic to the theory
and not because we were unable to make it better. This randomness means logical randomness, not statistical randomness
though they are related as we have seen in Sect.\ref{sec:not_compute}. Chaitin realized that logical randomness could be ubiquitous in Mathematics and started off the 
development of AIT in a form that can be considered  it as the natural evolution of the work by Turing.

G\"{o}del's theorem can be traced back to the `liar's paradox' \eqref{liar2} while Chatin's halting probability
is related to the `Berry's paradox':
\begin{equation}
\begin{array}{cc}
\text{`The smallest positive integer} \\
\text{not definable in under eleven words'}.
\label{berry}
\end{array}
\end{equation}
In principle, that proposition defines a certain positive integer since the set of words is finite while the set of integers is infinite. 
However, as that proposition has only ten words, it cannot be defined by that \eqref{berry}. This is the paradox. A similar situation arises in the definition of algorithmic complexity
\eqref{complexity}: if algorithmic complexity were computable by a TM, then similar paradoxes to \eqref{berry} would appear. Berry's paradox was formulated by
B. Russell inspired by a librarian at Oxford whose name was G.G. Berry. Chaitin explains that he wanted to show G\"{o}del in 1974 how he could prove the incompleteness 
theorem using Berry's paradox instead of liar's paradox \eqref{liar2}, but Chaitin was not able to meet  G\"{o}del.

He introduced the $\Omega$ number: the halting probability of a Turing machines \eqref{omega}. It is a natural example of a random infinite
sequence of bits. Besides providing a connection with the work of
Turing, $\Omega$ makes randomness in Mathematics more concrete and more believable. Chaitin has shown that this logical randomness
is at the very heart of pure Mathematics: provable theorems are islands surrounded by vast oceans of unprovable truths.

\subsection{David E. Deutsch}

David Deutsch culminated the formulation of a quantum computer in a way that
it is a well-established extension of the work by Turing into the quantum world.
R.P. Feynman gave fundamental steps prior to him,  as well as P. Benioff.
A precise definition of a quantum TM and its functioning  can be found in Galindo and Martin-Delgado \cite{rmp}.
Deutsch reformulated the Church-Turing thesis into a version usually called the Church-Turing-Deutsch principle:

``Every finitely realizable physical system can be perfectly simulated by a
universal model computing machine operating by finite means."

This is a farther extension of the Turing hypothesis into the physical world.

Quantum versions of algorithmic complexity, Sect.\ref{sec:not_compute}, has been formulated \cite{vitanyi_00,berthiaume_et_al_00,gacs_01,mora_briegel_05,mora_briegel_04,mora_briegel_06}, as well as  quantum versions of the $\Omega$ number 
\cite{svozil_95,svozil_95b}.

%%%%%%%%%%%%%%%%%%%%%%%%%%%%%%%%%%%%%%
%%%%%%%%%%%%%%%%%%%%%%%%%%%%%%%%%%%%%%
\section{Notions and definitions of Complexity}
\label{sec:complexity}
%%%%%%%%%%%%%%%%%%%%%%%%%%%%%%%%%%%%%%
%%%%%%%%%%%%%%%%%%%%%%%%%%%%%%%%%%%%%%

Complexity is a word, a password, that has proliferated in a large number of 
scientific disciplines: ...
Most of the times, its use is rather vague, volatile and qualitative.
After Turing, it is important to realize that a rigorous, mathematical definition of
complexity can be given and made quantifiable.

A very primitive and inefficient way to assess complexity in Mathematics is
to define it in terms of how long or difficult is to write the equations of a given theory.
Naive as it may look, its use is very extended in the scientific community. 
This is not appropriate since this notion is very dependent on the language we
use to write equations, and this may change over the times.
A proper definition of complexity calls for something more intrinsic.

If we want to quantify the complexity of a theory or discipline, we must seek how it relates to the
experimental data that it wants to explain. Thus, we consider the pair formed by a given  theory and its experimental data,  and map it into
another pair  which is a program that produces a certain  output: 
\begin{equation}
\fC: (\text{theory}, \text{data}) \longmapsto (\text{program}, \text{output}).
\label{mapping}
\end{equation}
This latter pair is related to a computer that takes the program and
finds the output.  We can call this a computational mapping $\fC$.
With this mapping, now we can apply complexity theory from computer science 
in order to find the complexity of a certain theory or discipline. This is an information-theoretic approach
to study  complexity  by using Turing's ideas in order to make things more precise.

In Information Theory (IT), there are two major notions of complexity: algorithmic complexity and computational complexity.

\noindent {\bf Algorithmic Complexity}:  it cares about the program-size complexity, i.e.,  bits of information regardless
the runtime of a computer, following the ideas of how a TM works. We have explained it in detail in Sect.\ref{sec:not_compute}.

This notion of complexity has no practical applications per se. It is very useful to study the fundamentals of Mathematics and its foundations. 

Although algorithmic complexity is rather conceptual, it may be also very inspiring in practical cases. There is an example that captures 
the essence of this complexity: the language used for storing image files. There are two basic procedures: using bitmap graphics or vectorial graphics.
The former corresponds to using all the bits of a given image and store them by brute force. The latter is more elegant since it tries to store the formula
that generates a certain graphics. This is more efficient and versatile since it preserves the image under change of scale.

A recent new development by Chaitin is to use AIT concepts and tools in order to give a mathematical proof of Darwin evolution theory \cite{chaitin_evo_11}.
With quantum versions of AIT, like new quantum $\Omega$ numbers, it is possible to study quantum effects in the theory of evolution \cite{martin_delgado_evo_11}.

For more practical purposes, the notion of computational complexity is preferred. Once a problem is declared computable, then we need to know if we can 
comput it efficiently or we can not. This leads to the notion of computational complexity.

\noindent {\bf Computational Complexity}: it evaluates the resources needed by a computer to solve a problem and how they scale with the typical size of the problem.
Time complexity refers to how many steps are needed to solve a problem. Space complexity refers to how much memory is needed to solve the problem.

Many computational tasks can be decomposed in simpler parts called decision problems.

\noindent {\bf Decision Problem}: it is a problem defined by an algorithm stated as a question whose answer is yes or not, equivalently, $1$ or $0$.
For instance: `Is $N$ a prime number?', and the like. Recall that we know from Sect.\ref{sec:compute}  that a Turing machine $T$ is the formal definition of an algorithm.
The TM associated to a decision problem is an application $T: \fX \rightarrow \{0,1\}$. 
Other important problems like `search' or `optimization' can be decomposed into decision problems.
Now, with the notion of a TM we can define precisely time complexity and space complexity.

\noindent {\bf Time Complexity}: Given a decision problem characterized by a TM $T$, it is the number of steps $t(N)$ that the TM takes before it halts and solves the problem. $N$ represents
the size of the input. One is normally interested in the study on the scaling of $t(N)$ for large $N$, or finding good upper bounds. Donald Knuth is an example of groundbreaking 
work on the analysis and performance of algorithms \cite{knuth_00}.

\noindent {\bf Space Complexity}:  Given a decision problem characterized by a TM $T$, it is the number of squares $s(N)$ of the work tape scanned by the TM before it halts.
Similar considerations apply as for time complexity.

It is very convenient to arrange sets of problems with the same complexity behavior into complexity classes.

\noindent {\bf Complexity Class}: is a set of decision problems that share the same type of time or space complexity according to some condition that is imposed on the problem,
which defines the class itself.

The most important class is the one that defines theoretically what an efficient algorithm is. This is the class P. 

\noindent {\bf P} : it is the class of decision problems that are solvable in polynomial time. The time of the algorithm, or associated TM, is bounded as $t(N)\leq c N^k$, for certain
$c\in \R$, $k\in \N$. The real constant $c$ is called the overhead of the algorithm, and it is convenient that the integer $k$ be the lowest possible.
Arithmetic operations like adding or multiplying, or the Gauss elimination method for solving linear equations are examples of algorithms in P.

\noindent {\bf PSPACE} : it is the class of decision problems that are solvable in polynomial memory space. Thus, the space of the algorithm, or associated TM, is bounded as
$s(N)\leq c N^k$, for certain $c\in \R$, $k\in \N$.

The class P is theoretically a natural choice of what an efficient algorithm is. The reason is for it is closed under operations that arise naturally in computation, like sum, product 
or composition of polynomials that are again polynomials. On the contrary, examples of inefficient algorithms are packed in the class EXP.

\noindent {\bf EXP} : it is the class of decision problems that takes an exponential time to solve them, $t(N)\leq {\rm e}^{p(N)}$, for some polynomial $p$.
For example, trial division to determine whether $N$ is a prime number is in EXP, and many brute force algorithms. 

A central problem in solving problems in computer science is the difference between finding a solution to a problem and verifying that a certain instance is a solution of the problem.
For instance, the decision problem `is $N$ a composite number?' is very difficult to solve for arbitrary $N$. However, if we are given a solution to this problem, say $M$, then
verifying this instance is a matter of division and this takes polynomial time. In this case, there are also polynomial algorithms to check whether $N$ is composite, but not for
finding its prime factors.
The general case can be casted in the form of a complexity class.

\noindent {\bf NP} : is the set of decision problems whose associated TM $T: \fX\times \fX \rightarrow \{0,1\}$ is in class P. $T(x,y)$ verifies whether the problem defined by
the bit-string $x$ once an instance $y$ is supplied. Additionally, the length of the verifier $y$ must be polynomially bounded: $|y|\leq p(|x|)$. 

With the advent of quantum Turing machines, the field of computational complexity has been revolutionized and enriched. New complexity classes can be defined substituting
the classical TM by a quantum version. For instance, the natural version of the class P for quantum computers is called BQP, for the class of bounded quantum polynomial problems.
Scott Aaronson has done systematic studies of a huge number of both classical and quantum complexity classes \cite{aronson}. Quantum Turing machines can also be generated
by sets of quantum gates \cite{yao_93}, what is known as the quantum circuit model.
Interestingly, it is possible to study the quantum
complexity of many statistical classical systems when simulated on a quantum computer and still find open problems \cite{vandenNest_09,delasCuevas_11}.

An example of complexity class relationship is $\text{P} \neq \text{EXP}$. Another is $\text{P} \subset \text{NP}$ and  $\text{NP} \subset \text{EXP}$.

\noindent {\bf P vs. NP Problem} : Is $\text{P} \neq \text{NP}$?

This is considered the central problem in computational complexity, and in computer science in general. Behind this question is whether computational creativity can be automated
or not. Thus, at first it looks like the natural answer to this problem is yes. 
However, there are neither proofs that  $\text{P} \neq \text{NP}$ or $\text{P} = \text{NP}$.

There is a third way to approach this problem. Notice that this problem is considered as a problem in complexity theory, not on computability.
However, this is not the case. True as it is that deciding whether a problem is either P or NP is a complexity problem, the P vs. NP problem is
equivalent to construct a mechanical procedure to decide whether it is true or false, and this is a problem on computability. Therefore, we have
to face also the possibility that it is non-computable. This means that it would be an irreducible axiom that one may or may not add to his theory
of computer science and go on to produce different types of theories, both equally valid and sensitive. Thus, if this third-way were true, then the
natural choice  $\text{P} \neq \text{NP}$ would be like Euclidean geometry, while the non-natural choice  $\text{P} = \text{NP}$ would be like 
non-Euclidean geometry. But this is also a conjecture.

There is not accepted definition of what a complex system is. Qualitatively, it is usually referred to a system compressed of various parts, usually many, 
such that they are interconnected somehow up to a certain degree, and the behavior of the whole system cannot be anticipated from the behavior of its individual parts.
Remarkably, this is precisely the situation that we basically have with a TMs working with simple binary system given rise to both computable and non-computable behaviors,
Sect.\ref{sec:compute}, \ref{sec:not_compute}. Thus, when the computational mapping \eqref{mapping} can be applied to a certain system, arbitrary as it may be, we may give a 
sufficient criterion for having complex behavior by appealing to the notion of hard problem:

\noindent {\bf NP Hard Problem}: when some problem, not necessarily in NP,  can be solved by an algorithm that can be
reduced to one capable of  solving any problem in NP, then it is called NP-hard.
A problem that is both NP and NP-hard is called NP-complete.

When some problem can be solved by an algorithm that can be
reduced to one that can solve any problem in NP, then it is called NP-hard.
A problem that is both NP and NP-hard is called NP-complete. These problems are at least as hard as the hardest problems in NP. 
Examples of NP hard problems are the `subset sum problem' and the `traveling salesman'. They both are also NP complete.
If $\text{P}\neq \text{NP}$, then $\text{NP}\neq \text{NP Hard}$, otherwise, they are equal.

%%%%%%%%%%%%%%%%%%%%%%%%%%%%%%%%%%%%%%
%%%%%%%%%%%%%%%%%%%%%%%%%%%%%%%%%%%%%%
\section{Some Applications}
\label{sec:applications}
%%%%%%%%%%%%%%%%%%%%%%%%%%%%%%%%%%%%%%
%%%%%%%%%%%%%%%%%%%%%%%%%%%%%%%%%%%%%%

%%%%%%%%%%%%%%%%%%%%%%%%%%%%%%%%%%%%%%
%%%%%%%%%%%%%%%%%%%%%%%%%%%%%%%%%%%%%%
\subsection{A Practitioner's Critique to Complexity Class P}
%%%%%%%%%%%%%%%%%%%%%%%%%%%%%%%%%%%%%%
%%%%%%%%%%%%%%%%%%%%%%%%%%%%%%%%%%%%%%

The notion of an efficient algorithm is defined by means of class P as explained in Sect.\ref{sec:complexity}.
There we saw that it is a good theoretical definition for this class P is closed under natural
operations that occur in computations. However, theoretically well-sounded as it may be, 
it runs into problems when dealing with practical cases and real computers. For instance,
an algorithm with a time complexity growing like $t(N) \sim N^{100}$ would never catch the interest
of any programmer. It would be good
to complement that notion of theoretical efficiency with another of `practical efficiency'.

Let us consider the following practical situation. We are given an algorithm whose time
complexity is in P as it grows like:
\begin{equation}
t(N)=c' N^k,
\label{real_time}
\end{equation}
where $t$ is now  the real clock-time taken by the computer to achieve the solution of a given problem
whose size is characterized by $N$. The integer $k$ is fixed by the time complexity of the algorithm,
and the constant $c'$ takes into account the conversion between theoretical time-steps and real time.
With the real computer we may have been able to obtain a certain set of points, simulation data: 
\begin{equation}
{\cal D}=\{t(1),t(2),\ldots,t(N_{\text{max}})\},
\label{data}
\end{equation}
up to a maximum achievable size $N_{\text{max}}$, which will depend on the technological resources available
when obtaining the data \eqref{data}. It may so happen, and it is currently the case, that the set of data is not enough to
discover a law we are searching for. This is another version of the situation thought by Leibniz in Sect.\ref{sec:precursors}.
Thus, we need a bigger value of $N_{\text{max}}$, but we are limited by the technological resources of our time, i.e., the time of the data \eqref{data}.
In order to assess how good the time complexity  \eqref{real_time} is, we need to compare with the estimated improvement of the technological resources.
An example of this is Moore's law for computers \cite{moore_65}. Following this, we may have found that our technology to build real computers behave as another power law
with respect to the minimum size $\ell_{\text{min}}$ of the computer chips that run the computations. Thus, the smaller the size the faster the computer:
\begin{equation}
t(N_{\text{max}})=c^{\prime\prime} \ell_{\text{min}}^{-\alpha},
\label{practicall_time}
\end{equation}
where $c^{\prime\prime}$ is a constant and $\alpha$ a scaling exponent known experimentally.

In order to discover the law, we need to increase the maximum current size $N_{\text{max}}$ by a certain factor $f>1$, such that the set of data up to $f N_{\text{max}}$
is now enough to determine the pattern. The question in turn is how much we need to improve our technology in order to achieve this. Thus, we can derive a sort of 
uncertainty relation between $N_{\text{max}}$ and $\ell_{\text{min}}$:
\begin{equation}
N_{\text{max}}^k \ell_{\text{min}}^\alpha = c,
\label{practical}
\end{equation}
with $c:=c^{\prime\prime}/c^{\prime}$ a fixed constant. The integer exponent $k$ is fixed by the class P of the algorithm and we want to know how to improve $N_{\text{max}}$
depending on the relative value of $\alpha$ w.r.t. $k$. Thus, we have $N_{\text{max}} = \text{const}/\ell_{\text{min}}^{\frac{\alpha}{k}}$. A possible situation could be that $k=\alpha$, then
a linear decrease in the chip technology will yield an increase in the maximum size. A better situation is when $k \ll \alpha$ since then the improvement will be over previous pay off.
However, the worst situation occurs when $k\gg \alpha$. In the limit case of $k\rightarrow \infty$, the maximum size would be insensitive to any technological improvement.

Therefore, a practical criterion for the class P is to compare the integer $k$ with the technological scaling exponent $\alpha$, i.e. $k$  vs.  $\alpha$, rather than the more theoretical
criterion of comparing  $k$  vs.  $\infty$.

Another important practical case we may face is the existence of technological barriers. For example, nowadays the computer technology has reached the size of the nanometers. Suppose we  have a certain set of data like \eqref{data} obtained with a class P algorithm, but we need to increase the maximum size by a factor $f>1$ such that then we need to
go down well beyond the size of amgstroms. Then, for those smaller sizes, the computer leaves the classical behavior and enter the realm of quantum mechanics, so that we may
well need a quantum computer to expand the range of data and be able to find our law.

%%%%%%%%%%%%%%%%%%%%%%%%%%%%%%%%%%%%%%
%%%%%%%%%%%%%%%%%%%%%%%%%%%%%%%%%%%%%%
\subsection{On the Halting Problem in Chess}
%%%%%%%%%%%%%%%%%%%%%%%%%%%%%%%%%%%%%%
%%%%%%%%%%%%%%%%%%%%%%%%%%%%%%%%%%%%%%

The halting problem has many implications as we have shown. It is a concept
of practical use in games, specially in advanced games like chess.  There, it is important
to make sure that the rules of the game (axioms) will ensure that the a match will
terminate. Until 1929, players were not aware that the set of rules known by then, allowed
to produce never-ending chess matches. In that year, Max Euwe, a mathematician later to become
the fifth world chess champion of modern history 1935-37 \cite{turing_years}, settled 
the question by rediscovering the Thue-Marston \cite{thue-marston} sequence and its cube-free property.

In binary language, the Thue-Marston sequence is defined by the following generating moves:
\begin{equation}
\begin{matrix}
t_0=0;\\
0 \mapsto 01; \  1 \mapsto 10.
\end{matrix}
\end{equation}
For instance, the first elements of the sequence are
\begin{equation}
\begin{matrix}
t_0 = 0, \\
t_1=01,\\
t_2=0110,\\
t_3= 01101001;\\
t_4= 0110100110010110.
\end{matrix}
\end{equation}
 An element of a sequence is cube-free if it contains no subsequence of the form $ppp$, where $p$ is a finite non-empty element.

A chess match is divided in three parts: opening, middle-game and final. The final part is characterized by the presence of very few
pieces on the board as compared to the opening. Thus, a theory of the final in chess has been developed to great extent: its complexity gets reduced. It had been known that
repetition of movements, a loop, may happen in certain situations. Rules were established to declare a draw when repetition of moves become endless.
Euwe \cite{euwe_29} used the cubic-free property of the Thue-Marston sequence to show how to circumvent those rules. Thus, new rules had to be added to the game.
This is another instance of how axioms, i.e. rules, may be changed a posteriori depending on the type of theory we want to have.

When Bobby Fisher was an active chess player, he would say ``Gods put the middle game after the opening", meaning that the complexity of the middle game
was so high that it was unknown territory, where written manuals for openings were useless, and he would feel at his best. After retirement, in the 1980's Fisher sent a warning
call saying that chess was becoming too technical, mechanical and with little room for creativity. He proposed to change the rules of the opening somehow, 
interestingly enough, introducing some randomness in chess. In particular, by randomizing the starting position of the main pieces in the first row of each player side.
And this happened way before a computer, Deep Blue,  defeated the World Chess Champion G. Kasparov in 1997.  Many people thought this to be unbelievable before year 2000.
This does not mean that computers are more intelligent than us, or intelligent at all. It means that their brute force of calculation is stronger than ours at playing chess.

%%%%%%%%%%%%%%%%%%%%%%%%%%%%%%%%%%%%%%
%%%%%%%%%%%%%%%%%%%%%%%%%%%%%%%%%%%%%%
\subsection{Divertimento: On the Complexity of Music}
%%%%%%%%%%%%%%%%%%%%%%%%%%%%%%%%%%%%%%
%%%%%%%%%%%%%%%%%%%%%%%%%%%%%%%%%%%%%%

Mozart composed many divertimentos, a musical form very common in the Classical era
prior to the success of the sonata form by Haydn. We may produce a divertimento playing
with Turing's ideas in music.

Music is more than a language, but as long as it is a language, we can apply
Turing's results to it and prove some amusing results which for music theorists
may be surprising, specially by the fact that they can be proved mathematically.

\noindent i/ There is an endless number of different musical compositions.% (from Cantor's diagonal method)

\noindent ii/ There are musical compositions that cannot be composed. % (from Turing's halting problem)

Statement i/ implies that musical creativity is infinity, for sure, while ii/ means that, nevertheless,  it  also has some limits.

To proof i/ we use a code  such that the music symbols  and rules of composition are encoded with a 
given alphabet ${\cal A}$. This can be binary for instance. Then, we use the same code alphabet
to label all known compositions. This can be done by lexicographic order, forming a table like \eqref{table}.
Now, applying  the diagonal method we obtain another composition which is certain to be new.
Though it is unlikely that these  mathematical type of compositions would have pleased Mozart and Haydn, 
it may produce a different reaction in B. Bart\'ok, A. Sch\"{o}ngberg, J. Cage, G. Ligeti, K. Stockhausen, I. Xenakis, P. Boulez, C. Halffter...
Nevertheless, what is remarkable, and unimaginable before Turing, is that a computer could be of help as a composition machine as they are used nowadays.

To proof ii/ we realize that each music composition is like a TM. Thus, it may or may not halt. For instance, we can produce simple scores that
repeat themselves forever. Accepting this proviso about endless compositions is essential.
Suppose now that we want to write a music composition that with our language be equivalent to a program that finds
when any other musical composition will ever halt. Then, that score is impossible to be written. 

In the beginning of XX century, Arnold Sch\"{o}ngberg evaluated the situation of classical music and judged that the tonal system based on
major and minor scales, Greek modes etc. was absolutely worn out. Subjective as this may be, he went on to search for new composition systems
by relaxing the rigidities of the old system. For instance, allowing all tones in a dodecaphonic scale to play the same role, without dominant or tonic tones.
This produce non-tonal systems like the twelve-tone method  and many others to follow, even by introducing random methods and  other tools from Mathematics,
like set theory.
Again, that situation arose because creativity was judged to be exhausted, and a change or extension of axioms was proposed instead, leading to controversy.
Nevertheless, controversy is unavoidable here since music is more than a language and personal taste plays a major role.

%%%%%%%%%%%%%%%%%%%%%%%%%%%%%%%%%%%%%%
%%%%%%%%%%%%%%%%%%%%%%%%%%%%%%%%%%%%%%
\section{Conclusions}
\label{sec:conclusions}
%%%%%%%%%%%%%%%%%%%%%%%%%%%%%%%%%%%%%%
%%%%%%%%%%%%%%%%%%%%%%%%%%%%%%%%%%%%%%

Turing has revolutionized the fundamental roots of what we understand by scientific knowledge,
and will continue to do so as many applications of his works will come up. At the same time, his scientific work still lacks the 
recognition that it deserves in his own field of Mathematics. As he also founded modern computer science, recognition
came first mainly from Engineering and Physics.

The part of Turing's 1936 paper \cite{turing_36} devoted to computable numbers has given rise to the development
of the whole computer technology. This is having a gigantic impact in our culture. The other part of Turing's paper devoted
to the solution of Hilbert's tenth problem, as a consequence of the previous one, has helped us to deepen our knowledge about
scientific knowledge itself. This is best exemplified by the work of Chaitin, who has formalized what is knowable and unknowable
based on Turing's work, and extending G\"{o}del's results in a more systematic and accessible way. His conclusion is rather shocking
since it implies that logical randomness is common even in Mathematics.

There is a parallelism between intrinsic randomness in Mathematics and in Physics, and we can learn from it.
In Physics it appeared in 1920's in Quantum Mechanics, and also produced a shocking revolution that removed the holy grail
of classical Physics, determinism, from its central status it had been enjoying. Nowadays, Quantum Mechanics is a successful theory
and has been accepted both logically and due to its unprecedented accurate experimental results.  In Mathematics, logical randomness appeared
in 1930's and it will so happen that will become accepted.

After the work of G\"{o}del, Turing and Chaitin it is certain that a TOE of Mathematics is impossible. But, what about Physics? Inasmuch as
Physics inherits the language of Mathematics to express its laws and works out its consequences, we may  immediately deduce that the same applies
to Physics and there is no TOE for it. However, Physics is more than a language and the ultimate word relies on experience, on the natural law. 
Our physical knowledge is like a window in an energy scale, ranging from some point in the infrared to some point in the ultraviolet, i.e., large distance
scales to small distance scales. From this finite window scale we may bet on two possibilities: i/ that no TOE of Physics exists, since as we enlarge the
energy window we will get new laws of Physics that were not anticipated; ii/ that a TOE of Physics do exists and from our current window of knowledge,
or probably a better one, we can deduce the whole range of physical laws in the entire energy scale, i.e., Physics would be finite and closed as a source
of knowledge. Following Turing's work, I believe that option i/ is the correct one, and experience will tell us. The non-existence of a TOE in Physics is good 
news for creativity in contrast to reductionism.

Is it true that true randomness is only quantum? The $\Omega$ number is a real number whose binary expansion yields bits of information
that are true for no reason, they have no structure or pattern, it is incompressible and its bits totally random as Chaitin has shown.

The following question  may help to face this outcome situation not so dramatically.
How can it be that Natural Sciences like Physics, Mathematics, etc have become so successful
if we live in a world plagued by intrinsic randomness? A clue to this question is to take the example
of what type of real numbers are employed in successful theories. We will see that we always have
real numbers like $\sqrt{2}$, $\pi$, ${\rm e}$, etc. Although they are irrational with an infinite number of 
decimals, we have very short algorithms that generate that series of decimals very efficiently.
I.e., they are actually maximally compressible numbers. 

This fact can be extrapolated to the whole structure of successful  theories of Nature: they are very simple,
they can be compressed, reduced to a simple set of axioms or laws of Nature. The rest of the universe that
remains unknown is due in part because it is not compressible and we live in a small region of the whole space
of theories or knowledge. We may divide our 'sphere of knowledge' into three parts: i/ current science (known); ii/ future science (to be known);
and iii/ unknowable or irreducible.

%Figure: the 'sphere of knowledge' divided in I: current science (known); II: to be known; III unknowable or irreducible.

Physicists are willing to find and adopt new physical principles, laws that expand their knowledge of the universe. Mathematicians, standard and formal ones, tend to stick rigidly to axioms and not to modify them. They should adopt a more experimental attitude. With Turing, the fields of Mathematics and Physics become more unified.

\begin{acknowledgments}

M.A.M.-D. thanks the Spanish MICINN grant FIS2009-10061,
CAM research consortium QUITEMAD S2009-ESP-1594, European Commission
PICC: FP7 2007-2013, Grant No.~249958, UCM-BS grant GICC-910758.

\end{acknowledgments}

%%%%%%%%%%%%%%%%%%%%%%%%%%%%%%%%%%%%%%%%%%%%%%%%%%%%%%%%%%%%%%%%%%%%%%%%%%%%%%
%\begin{references}


\begin{thebibliography}{99}
\bibitem{turing_years} These are years when Turing was producing his seminal works.
\end{thebibliography}
\end{document}